\documentclass[12pt]{iopart}

\usepackage{iopams}
\usepackage{amssymb}
\expandafter\let\csname equation*\endcsname\relax
\expandafter\let\csname endequation*\endcsname\relax
\usepackage{amsmath}
\usepackage{epsfig}
\usepackage[english]{babel}
\usepackage{hyperref}
\usepackage{lineno}
\newcommand{\BE}{\begin{equation}}
\newcommand{\EE}{\end{equation}}
\newcommand{\BA}{\begin{eqnarray}}
\newcommand{\EA}{\end{eqnarray}}
\newcommand{\cG}{\mathcal{G}}
\newcommand{\cL}{\mathcal{L}}
\newcommand{\cF}{\mathcal{F}}

\emergencystretch=\maxdimen
\begin{document}

\title[Pulsed interactions unify pattern-forming reaction-diffusion and nonlocal models]{Pulsed interactions unify reaction-diffusion and spatial nonlocal models for biological pattern formation}

\author{Eduardo H.  Colombo}
\address{Center for Advanced Systems Understanding, Untermarkt 20, 02826 G\"{o}rlitz\looseness=-1}
\address{Helmholtz-Zentrum Dresden-Rossendorf, Bautzner Landstra{\ss}e 400, 01328 Dresden\looseness=-1}
\ead{e.colombo@hzdr.de}

\author{Ricardo Martinez-Garcia}
\address{Center for Advanced Systems Understanding, Untermarkt 20, 02826 G\"{o}rlitz\looseness=-1}
\address{Helmholtz-Zentrum Dresden-Rossendorf, Bautzner Landstra{\ss}e 400, 01328 Dresden\looseness=-1}
\address{ICTP - South American Institute for Fundamental Research \& Instituto de F\'{\i}sica Te\'{o}rica,
Universidade Estadual Paulista - UNESP, S\~{a}o Paulo, SP,
Brazil.}
\ead{r.martinez-garcia@hzdr.de }

\author{Justin M. Calabrese}
\address{Center for Advanced Systems Understanding, Untermarkt 20, 02826 G\"{o}rlitz\looseness=-1}
\address{Helmholtz-Zentrum Dresden-Rossendorf, Bautzner Landstra{\ss}e 400, 01328 Dresden\looseness=-1}
\address{Department of Ecological Modelling, Helmholtz Centre for Environmental Research - UFZ, Leipzig, Germany}
\address{Department of Biology, University of Maryland, College Park, MD, USA.}
\ead{j.calabrese@hzdr.de}

\author{Crist\'obal L\'opez}
\address{Instituto de F\'{\i}sica Interdisciplinar y
Sistemas Complejos (IFISC), CSIC-UIB, Campus Universitat Illes
Balears, 07122, Palma de Mallorca, Spain }
\ead{clopez@ifisc.uib-csic.es}

\author{Emilio Hern\'andez-Garc\'{\i}a}
\address{Instituto de F\'{\i}sica Interdisciplinar y
Sistemas Complejos (IFISC), CSIC-UIB, Campus Universitat Illes
Balears, 07122, Palma de Mallorca, Spain }
\ead{emilio@ifisc.uib-csic.es}

\begin{abstract}
The emergence of a spatially-organized population distribution depends on the dynamics of the population and
mediators of interaction (activators and inhibitors).
Two broad classes of models have been used to investigate when and how self-organization is triggered, namely, reaction-diffusion and spatially nonlocal models. Nevertheless, these models implicitly assume smooth
propagation scenarios, neglecting that individuals many times interact by exchanging short and abrupt pulses of the mediating substance. A recently proposed framework advances in the direction of properly accounting for these short-scale fluctuations by applying a coarse-graining procedure on the pulse dynamics.
In this paper, we generalize the coarse-graining procedure and apply the extended formalism to new scenarios
in which mediators influence individuals'
reproductive success or their motility.  We
show that, in the slow- and fast-mediator limits, pulsed interactions recover, respectively, the reaction-diffusion and nonlocal models, providing a mechanistic connection between them. Furthermore, at each limit, the spatial stability condition is qualitatively different, leading to a timescale-induced transition where spatial patterns emerge as mediator dynamics becomes sufficiently fast.
\end{abstract}



\section{Introduction}
Living organisms are capable of self-organizing to
create complex spatiotemporal patterns~\cite{camazine2020self,martinez2022spatial}.
Well-known examples of these patterns are
produced by plants~\cite{Klausmeier1999,von2001,Fernandez-Oto2014},
 mussels~\cite{koppel2005scale}, birds~\cite{vicsek,attanasi2014information}
 and bacteria~\cite{ben1990,tyson1999minimal}.
 In all these examples, the patterns are both  visually striking and functionally significant. Specifically, the patterns increase populations'
 resilience and robustness against variations in environmental conditions~\cite{liu_pattern_2014,bonachela_termite_2015,rietkerk_self-organized_2004,Martinez-Garcia2023} or population size \cite{Jorge2023}, and can also reduce predation risk and movement
 costs~\cite{parrish1999complexity}. Therefore, the study
 of self organization, beyond contributing to the fundamental
 understanding of emergent phenomena in living systems,
 is also relevant for understanding ecosystems'
 large-scale properties and can shed light on how to
 design and implement management strategies~\cite{levin1992problem}.

An overview of theoretical studies that investigated 
the mechanisms behind spatial self-organization identifies two
broad classes of models. First, \emph{reaction-diffusion}
models, strongly inspired by Turin's seminal
work~\cite{Turing1952a}, describe the dynamics of
the density distribution of a population, and of substances 
that govern inter-individual interactions, i.e. \emph{mediators}~\cite{rietkerk}. Mediators can be, for example, physicochemical substances (light, sound, toxins, nutrients) that are produced or consumed and have the potential to affect exposed individuals. Here, we will consider a reaction-diffusion (population-mediator) system consisting of a pair of partial differential equations, where the rates are local.
Second, in an alternative family of models the explicit description of the mediator dynamics is absent and
only the population dynamics is described, using
a single equation for the density of organisms. The impact
of the mediating substance on population dynamics appears as
an effective, spatially extended interaction
between organisms at distant locations, modeled with a phenomenological interaction kernel
(or influence function)~\cite{Fuentes2003,emilio1,Clerc2005,Pigolotti2007,lefever1997origin,sasaki1997clumped}. We refer to this class of models as \emph{nonlocal models} (also commonly known as kernel-based models).

The choice of which class to use depends on the
goal of the study and the level of knowledge
about the dynamics of the mediators. For example,
in the case of vegetation pattern formation,
models belonging to these two types have both
been used. Using the reaction-diffusion structure,
the dynamics of the vegetation biomass
and water is described explicitly by means
of two (or more) partial differential equations\cite{bastiaansen2018,meron2019}.
Alternatively, other studies have described
the same vegetation-water system using
a single spatial non-local partial differential
equation \cite{lefever1997origin,Martinez-Garcia2023}. For that, an interaction kernel
is used to describe how the net interaction between plants
(as a result of water-vegetation feedbacks)
changes in strength and sign with the
distance between them
(see \cite{Martinez-Garcia2023} for a review about the two approaches).

However, reaction-diffusion and non-local models
are defined already at a coarse level 
in which population and mediator
densities change slowly and smoothly with time.
These formulations therefore do
not guarantee that short-timescale
fluctuations are properly taken into account,
which might yield misleading predictions given the
variety of ways in which mediators
can be released and consumed by
organisms, thus affecting their propagation in space and time ~\cite{smith2003animal}.
In fact, mediators can be of different types and propagation rules vary widely, as organisms may use
acoustic~\cite{ricardo2013p}, visual~\cite{caro2017} or
chemical~\cite{ojalvo2018} signals to attract, repel,
slow down or speed up con- and heterospecifics
~\cite{zinati2022stochastic,grimaCurrent,grimaPRL,Lopez2004}.

\begin{figure}[h]
    \includegraphics[width=1\columnwidth]{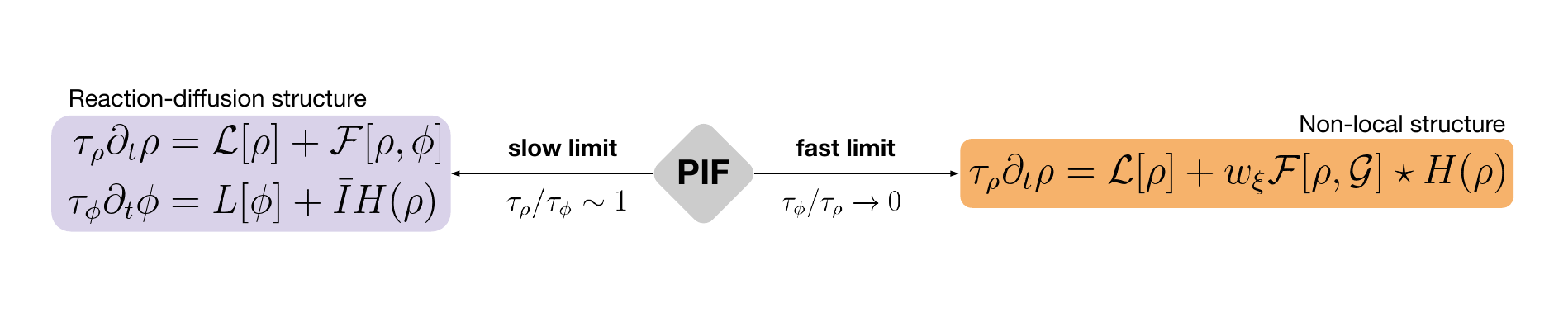}    \caption{\textbf{Pulsed Interaction Framework (PIF).} The framework considers the  dynamics of the density distribution of the focal population, $\rho(x,t)$, and an interaction mediator, $\phi(x,t)$. The dynamics is ruled by the functionals $\mathcal{L}$, $L$ and $\cF$, which are associated, respectively, with the population, mediator, and the coupling between them, and by the pulses of the mediator stochastically released by the population. In the slow-mediator dynamics limit, for which the mediator and population dynamics timescales are comparable ($\tau_\phi/\tau_\rho \to 1$), the framework recovers the structure of two partial differential equations, referred to as a reaction-diffusion system (with local rates). In the fast-mediator dynamics limit, the mediator timescale is much shorter than the one associated with population dynamics ($\tau_\phi/\tau_\rho \to 0$), and this leads to a single equation with nonlocal interactions, as it depends on $\cF[\rho,\cG] \star H(\rho)=\int dx' \cF\left[\rho,\cG(x-x')\right]
H(\rho(x',t))  $, where $\mathcal{G}$ is the influence function, also known as the interaction kernel (see Sec. \ref{sec:framework} for derivation and detailed discussion).}
    \label{fig:scheme}
\end{figure}

To address how the temporal variability in mediator
propagation dynamics can affect population
self organization, Colombo et al.~\cite{Colombo2023}
recently proposed a framework in which individuals attack
each other by releasing short toxic
pulses into the environment.
By properly coarsening the individual-level
dynamics, this study showed that the
toxin-field response to the pulsed releases can completely
change the spatial stability of the
population and, consequently,
the outcome of the population dynamics. In a limit of
slow-toxin dynamics,
the model recovers the
reaction-diffusion scenario~\cite{Turing1952a},
while in a fast toxin-dynamics limit,
an effective description emerges in terms of
a single equation for the population
dynamics with a spatially-extended
interaction that arises from toxin spatial propagation~\cite{Borgogno2009,murray2003}.

In this work, we generalize these results to account for other
different pathways of interaction found in nature by
constructing a general pulsed interaction framework (PIF; see
Fig.~\ref{fig:scheme}), which contributes to a general theory
for self-organized spatial pattern formation in living systems.
In particular, in addition to the previously considered case of
a toxin as mediator, we address here possible scenarios of
signal release that influence the mobility of individuals.
Mathematically, this extension means to take into account the
possibility, via a more general coupling functional
$\mathcal{F}$ (see Fig.~\ref{fig:scheme}), of terms that link
the mediator density, $\phi$, the population density, $\rho$,
and their respective derivatives. Previously, only the death
rate was influenced by mediator density, and this was done via
a linear coupling involving no gradients, $-\rho\phi$. In the
new formalism the mediator density can also affect motility
processes, and in an arbitrary nonlinear way. In addition to
that, a general density-dependent pulse rate is allowed, thus
generalising the linear pulse rate of \cite{Colombo2023}.

\section{A general framework for pulsed interactions}
\label{sec:framework}
 We model an ensemble of simple
organisms in a one-dimensional spatial domain, but extensions to higher-dimensional setups should be straightforward. 
Organisms move, reproduce, and release mediators
in the form of
pulses. These pulses can have either biochemical (pheromones, autoinducers, and signaling molecules)
 or physical (heat, sound, electricity
or light) origins, and can affect conspecific survival~\cite{foster2013,burt1943,caro2017}
or movement~\cite{grimaCurrent,miller2001quorum}. We describe
this scenario at the population level by the following  equations,
\begin{align}
\label{geneq1}
\tau_\rho\partial_t \rho &= \mathcal{L}[ \rho] +  \mathcal{F}[\rho,\phi] \, ,\\
\tau_\phi\partial_t \phi &= L[ \phi] + \xi_\rho\, ,
\label{geneq2}
\end{align}
where $\rho$ and $\phi$ are the population density and signal or mediator
density, respectively. $\mathcal{L}$ and $L$ describe the dynamics of the population and the mediator when not coupled with each other.
The use of square brackets as in $\cL[\rho]$ or $L[\phi]$ stresses that these quantities are not simple
functions of their arguments, but functionals that can depend also on spatial gradients
such as $\partial_x\rho$ or $\partial_x\phi$. $\cF$ gives the effect of the mediator $\phi$ on the population dynamics. Without loss of
generality we assume that $\cF[\rho, 0] = 0$. $\tau_\rho$ and $\tau_\phi$ explicitly set the timescales for the
population and mediator dynamics, respectively.

The mediator field is released by the population
according to a stochastic state-dependent
protocol, $\xi_\rho$, which depends on the
population density, thus providing a coupling from the population to the mediator dynamics. More specifically, we consider that individuals release the mediating substance
in the form of a set of pulses localized in space and of short duration $w_\xi$:
\begin{align}
\label{releases}
\xi_\rho(x,t) &= \sum_i I_0 \Pi(t-t_i,w_\xi)\delta(x-x_i)\, ,
\end{align}
with $I_0$ being the amplitude of the unit
square pulse, $\Pi(t,w_\xi)$, which starts at $t=0$ and lasts until $t=w_\xi$ (i.e. $\Pi(t,w_\xi)=1$ if $0<t<w_\xi$ and $\Pi(t,w_\xi)=0$ elsewhere). We assume that individuals trigger pulses in such a way that $\xi_\rho(x,t)$ is a spatiotemporal Poisson process with density-dependent rate $H(\rho(x,t))$, i.e. the probability that a pulse starts in a neighborhood of size $\Delta x$ and $\Delta t$ of the space-time location $(x_i,t_i)$ is given by $H(\rho(x_i,t_i))\Delta x \Delta t$. Thus, the probability density for a pulse to occur at location $x$ is $\textrm{pdf}_t(x)=H(\rho(x,t))/R_H(t)$, where $R_H(t) \equiv \int_{-\infty}^{+\infty}H(\rho(x,t))dx$. Also, the occurrence of pulses anywhere in the system is a renewal process with time-dependent rate $R_H(t)$. If $\rho$ changes slowly enough at the scale of the pulses, the time interval between two consecutive pulses occurring close to the time $t$ can be approximated by $\tau_\xi = 1/R_H(t)$.

A natural choice for the rate $H$
is $H(\rho)=\alpha \rho$, meaning that each individual in the population releases mediator pulses
at a constant rate $\alpha$ independently of the others. Another choice that will be used later is $H = \alpha (\rho/\bar\rho)^\beta$, for which the per-capita release rate
increases ($\beta>1$) or decreases ($\beta<1$) with local density, mimicking a kind of quorum sensing.

It should be stressed that, although we are talking about `individuals', our model in Eqs. (\ref{geneq1})-(\ref{geneq2}) describes the system as continuous fields and does not resolve these individuals. The only remnant of the discrete nature of the population is that releases of mediator occur at stochastic discrete space-time points.

There are four crucial timescales: $\tau_\rho$,
$\tau_\phi$, ${w_\xi}$ and $\tau_\xi$. We will focus on
cases in which pulse duration ($w_\xi$) is much shorter than inter-pulse time ($\tau_\xi$), which is itself much shorter than population
dispersal and other demographic processes, $w_\xi \ll
\tau_\xi \ll \tau_\rho$. This means that there is a timescale separation between the occurrence of interaction events and the consequences of those events on population dynamics. Note also that timescale $\tau_\xi$ is associated to the population dynamics (which triggers the pulses) and not to the mediator  dynamics, being the mediator-dynamics timescale given exclusively by $\tau_\phi$.

In the following we investigate how the system's spatial
stability changes as a function of the mediator-dynamics timescale,
$\tau_\phi$. We obtain coarse-grained descriptions for the
population-mediator system (\ref{geneq1})-(\ref{releases}) and the respective pattern-forming
conditions for a) the \emph{slow mediator-dynamics limit}, in
which the mediator dynamics is relatively slow, being comparable to
population dynamics timescales $\tau_\phi/ \tau_\rho \sim 1$, and thus $w_\xi, \tau_\xi \ll \tau_\phi$; and b)
\emph{fast mediator-dynamics limit}, when mediator response is the fastest of all the timescales, $\tau_\phi \ll w_\xi$, so that $\tau_\phi/\tau_\rho \to 0$.

\subsection{Slow mediator-dynamics limit}

When $\tau_\phi/\tau_\rho
\sim 1$ the inter-pulse time is much shorter than the
population and mediator timescales, $\tau_\xi \ll \tau_\phi,
\tau_\rho$.
When this occurs, the mediator field $\phi$ in Eq.~(\ref{geneq2})
effectively feels only the average of the pulses
triggered by the population, which are many
and occur too fast for $\phi$ to respond to them. Consequently, we
can replace $\xi_\rho$ by its average over small time
windows of duration $\Delta t$, with $ \tau_\xi \ll  \Delta t \ll  \tau_\phi$, and spatial intervals of size  $\Delta x$:
\begin{align}
\label{avgdef}
\langle \xi_\rho(x,t) \rangle \equiv
\frac{1}{\Delta t \Delta x} \int_{-\Delta t}^{0}\int_{-\frac{\Delta x}{2}}^{\frac{\Delta x}{2}}\xi_\rho(x+x',t+t') dx' dt' .
\end{align}

Using Eq. (\ref{releases}), $\langle \xi_\rho(x,t) \rangle  = (\Delta x\Delta t)^{-1}\sum_{i=1}^{n_\xi} I_0
w_\xi  =  I_0 w_\xi n_R/(\Delta x \Delta t)$, where
$n_R$ is the number of pulses that have occurred during the
focal space-time window. Noting that pulses are
independent events, $n_R$ follows a Poisson
distribution with mean $H(\rho(x,t))\Delta x \Delta t$.
As $\Delta t \gg \tau_\xi$, $n_R$ is also large and its
ratio of standard deviation to mean vanishes,
so that fluctuations can be neglected. Thus $n_R \approx
H(\rho(x,t))\Delta x \Delta t$. As a consequence, in the
slow mediator-dynamics limit,
\begin{equation}
    \langle \xi_\rho(x,t) \rangle
\approx I_0 {w_\xi}H(\rho(x,t))\, .
\end{equation}

The other terms in Eqs. (\ref{geneq1})-(\ref{geneq2}) can also be coarse-grained but, due to their slow response times, they remain constant and unaffected by the averaging: $\langle \rho
\rangle \approx \rho$, $\langle \phi \rangle \approx \phi$,
$\langle \cL  \rangle \approx \cL$, $\langle \cF \rangle \approx \cF$. Thus, the final form for our equations in this slow-mediator limit ($w_\xi\ll \tau_\xi\ll
\tau_\phi,\tau_\rho$) is
\begin{align}\notag
\tau_\rho\partial_t \rho &= \cL[ \rho] + \cF[ \rho,\phi]\, ,\\
\tau_\phi\partial_t \phi &= L[ \phi] + \bar I H(\rho)  \, ,
\ \ \textrm{with} \ \ \bar I \equiv   w_\xi I_0,
\label{sloweqs}
\end{align}
which is simply equivalent to the situation in which
the pulsed mediator releases $\xi_\rho$ are replaced by the mean value $\bar I H(\rho)$.

\subsection{Fast mediator-dynamics limit}
\label{sec:fastmediator}
In this limit the mediator dynamics is much faster than any other process: $\tau_\phi \ll w_\xi \ll \tau_\xi \ll \tau_\rho$ (in particular $\tau_\phi/\tau_\rho \to 0$). Therefore, during a pulse occurring at location $x_p$, the mediator field rapidly reaches a stationary profile, $\mathcal{G}(x-x_p)$ (we assume, as will be the case in the examples below, that such stationary profile actually exists; we also assume that the solution $\phi$ of Eq. (\ref{geneq2}) tends rapidly to $\phi=0$ when $\xi_\rho=0$). This mediator profile during the pulse can be obtained by solving
\begin{equation}
\label{G}
    L[\cG(x)]+ I_0\delta(x-x_p)=0 \ .
\end{equation}
For instance, if $L[\cG(x)] = (D\partial_{xx} - \gamma)\cG(x)$, $\mathcal{G} \propto e^{-|x-x_p|\sqrt{\gamma/D}}$ (see Eq.~(\ref{kernel}) for another example).

The conditions $\tau_\phi \ll w_\xi \ll \tau_\xi$ guarantee that
pulses are non-overlapping, so that
the solution of Eq. (\ref{geneq2}) can approximated by a sum of the induced concentration profiles
during pulses that occur at different locations and instants $(x_i,t_i)$:
\BE
\label{eq:phipulsed}
\phi(x,t) = \sum_i \cG(x-x_i)\Pi(t-t_i,w_\xi) \ .
\EE
We see that the response of the mediator field $\phi$ to the release pulses $\xi_\rho$ consists also of a sequence of pulses, but in contrast to $\xi_\rho$, the pulses in $\phi$ are spatially-extended, having a spatial structure ($\cG(x)$ instead of $\delta(x)$) determined by $L[\phi]$.

Once Eq. (\ref{geneq2}) has been solved, we can substitute this expression for $\phi$ in the population Eq. (\ref{geneq1}):
\BE
\label{eq:Ggeneq1}
\tau_\rho\partial_t \rho = \mathcal{L}[\rho] +  \mathcal{F}\left[\rho, \sum_i \cG(x-x_i)\Pi(t-t_i,w_\xi)\right] \ .
\EE

Because the dynamics of $\rho$ is much slower than the sequence of pulses, we can simplify the $\rho$ dynamics by performing, as in Eq. (\ref{avgdef}), an average over a small space-time cell. The terms depending only on $\rho$ will not be affected by this coarse-graining: $\langle \rho\rangle\approx\rho$, $\langle\cL[\rho]\rangle \approx \cL[\rho]$, but we should compute the average $\langle \cF\rangle$. To show how this is done, we first calculate the average for the simple case in which $\cF(\rho,\phi)=\phi$, with $\phi$ given by Eq. (\ref{eq:phipulsed}):
\BE
\left<\phi(x,t)\right> \equiv \frac{1}{\Delta t \Delta x}
\int_{-\Delta t}^0\int_{-\frac{\Delta x}{2}}^{\frac{\Delta
x}{2}} \sum_i \mathcal{G}(x+x'-x_i)\Pi(t+t'-t_i,w_\xi)
dt' dx' \ .
\EE
As before, the temporal interval $\Delta t$ should satisfy $\Delta t \ll
\tau_\rho$ but we will also assume that it is larger than other
microscopic time scales: $w_\xi, \tau_\xi \ll \Delta t$. In contrast to the slow case, here the spatial
coarse-graining is not really needed, since in space $\phi$ is a smooth function. Thus, we eliminate the spatial coarse-graining by taking the limit $\Delta x\rightarrow
0$. The remaining temporal integral acts on the indicator
function $\Pi$, selecting at any time $t$ only the
$m_R$ pulses that have occurred anywhere in the system during the interval $[t-\Delta t,t]$. Thus we have:
\BE
\langle \phi(x,t) \rangle \approx (\Delta t)^{-1}
w_\xi \sum_i^{m_R} \mathcal{G}(x-x_i) .
\EE
$m_R$ follows a Poisson distribution with mean $R_H(t) \Delta t$.
We have neglected the values of $t$ for which a pulse is only
partially contained in $\Delta t$. Indeed, such time intervals are very small and negligible at the population scale $\tau_\rho$ because of the condition
$w_\xi \ll \Delta t$.

As in
the calculation for the slow-mediator case, the condition $\tau_\xi
\ll \Delta t $ implies a large value of $m_R$ so that, by the law
of large numbers, its fluctuations can be neglected.
We can also use $m_R^{-1}\sum_i \mathcal{G}(x-x_i) \approx
\int dx' \mathcal{G}(x-x') \textrm{pdf}_t(x')$, with
$\textrm{pdf}_t(x)=H(\rho(x,t))/R_H(t)$ being the probability density
of the locations $x_i$. Combining these results we arrive at
\BE
\label{meanphi}
\left<\phi\right>\approx {w_\xi}  \int
\mathcal{G}(x-x') H(\rho(x',t)) dx' \equiv  {w_\xi}
[\mathcal{G}\ast H(\rho)]\, ,
\EE
where $\mathcal{G}\ast H(\rho) \equiv \int
\mathcal{G}(x-x') H(\rho(x',t)) dx'$.

After this simple case, we turn now to the calculation of $\langle \cF \rangle$ in the general situation. We first note that $\Pi(t-t_i,w_\xi)$ takes only the values 0 or 1, that we have assumed that $\cF[\rho,0]=0$, and that there is at most one pulse active at each time $t$. Thus, we have
\BE
\mathcal{F}\left[\rho, \sum_i \cG(x-x_i)\Pi(t-t_i,w_\xi)\right] = \sum_i\cF\left[\rho,\cG(x-x_i)\right]
\Pi(t'-t_i,w_\xi) \ ,
\EE
from which (only the temporal average is needed):
\BA
\label{eq:avgphi}
&&\langle \cF(\rho,\phi) \rangle = \frac{1}{\Delta t} \int_{-\Delta
t}^0 dt' \left[\sum_i \cF\left[\rho, \cG(x-x_i)\right]
\Pi(t'-t_i,w_\xi)\right] \nonumber \\
&=& \frac{w_\xi}{\Delta t} \sum_i^{m_R} \cF\left[\rho,\cG(x-x_i)\right] \approx
\frac{w_\xi}{\Delta t} m_R \int dx' \cF\left[\rho,\cG(x-x')\right]
\frac{H(\rho(x',t))}{R_H}  \nonumber \\
&\approx&  w_\xi \int dx' \cF\left[\rho,\cG(x-x')\right]
H(\rho(x',t)) \equiv w_\xi \cF[\rho,\cG] \star H(\rho) \ .
\EA
Note that we have introduced the symbol $\star$ to denote a
partial convolution in which only the second argument of $\cF$,
and not the first one, is integrated. This structure arises
naturally, as $H$ appears when averaging over the statistics of
mediator releases, which is the quantity in the second argument
of $\mathcal{F}$, and the averaging does not directly affect
the first argument, the population density, since it has a much
slower dynamics.

Summarizing, the fast mediator dynamics $(\tau_\phi \ll w_\xi \ll \tau_\xi \ll \tau_\rho)$ gives
\begin{align}
\label{fasteq}
\tau_\rho \partial_t \rho = \cL[\rho]+ w_\xi \cF[\rho, \cG] \star H(\rho)  \ .
\end{align}

Because of the slow behavior of the population density, the stochastic nature of the pulsed releases does not appear in the population description for the two limits considered, i.e. Eqs. (\ref{sloweqs}) and (\ref{fasteq}) are both deterministic descriptions. However, the mathematical structure of each limit is completely different, and it is natural to expect that their
spatial stability also differs. In the next sections,
we provide concrete examples that show how population
spatial organization changes when we go from slow to fast propagating mediators.

\section{Applications}
\label{sec:applications}
The following concrete examples will each focus
on mediators that affect either growth/death or mobility, inspiring possible applications of the framework in the biological and ecological contexts. Our aim is to show dynamics that lead to pattern formation when the pulsed character of the mediator releases is more evident (i.e. close to the fast-mediator limit) but for which spatial patterns disappear when mediator dynamics becomes much slower than the pulse rate and duration or, equivalently, when only the mean effect of the pulses is taken into account.

For definiteness, we will restrain ourselves to
a class of mediators that diffuse and decay with state-dependent rates \cite{Colombo2023},
\begin{equation}
\label{signal_nl_dynamics}
    L[\phi] = D_\phi\partial_x(\phi^{\nu-1}\partial_x \phi) - [\gamma \phi^{\mu-1}]\phi\, . \,
\end{equation}
so that Eq. (\ref{geneq2}) reads
\BE
\label{specificeq2}
\tau_\phi\partial_t\phi=D_\phi\partial_x(\phi^{\nu-1}\partial_x \phi) - \gamma \phi^\mu +\xi_\rho \, ,
\EE
where $\mu,\nu>0$ are positive exponents that control the level of nonlinearity of the diffusion-decay process~\cite{Tsallis1996}.
This class of models embodies a wide range of situations
when diffusion and dissipation processes are state-dependent, including cases where both the diffusion and decay rate are
constant ($\nu,\mu = 1$), decreasing
($\nu,\mu < 1$), or increasing ($\nu,\mu > 1$) with mediator density.
Density-dependent diffusion can represent the population-level  impact of many physical~\cite{Cates2010} and biological~\cite{murray2003} factors acting on individuals' motility. For example, density-dependent diffusion can arise due to the spatial structure of the medium, as observed in propagation of liquids, gases and bacteria through porous media ($\nu>1$)~\cite{porous,bacporous} as well as via biological interactions (e.g. attracting and repelling forces) between the spreading entities, which can lead to $\nu>1$ or $\nu<1$~\cite{murray2002,murray2003}. Similarly, a density-dependent decay rate may arise due to medium porosity~\cite{bear2013} or via biological interactions, which can introduce Allee-like ($\nu>1$) and harmful crowding ($\nu<1$) effects at low densities that make the lifespans of biological entities shorter or longer, respectively~\cite{murray2002,okubo2013,allee,Colombo2018}.

\begin{center}
    \begin{figure}[h]
        \centering
        \hspace{2cm}        \includegraphics[width=0.8\columnwidth]{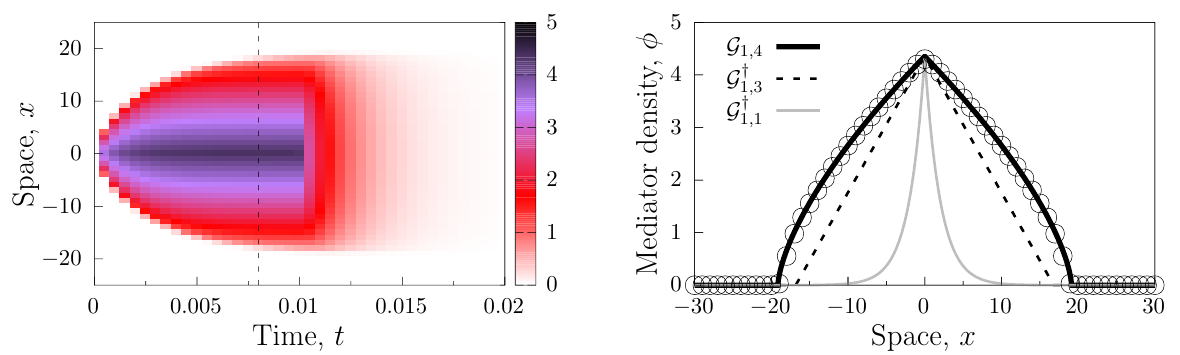}
        \caption{\textbf{Mediator profile during pulse}. A single pulse is introduced at the origin leading to the growth and spread of mediators. Left panel shows in color the values of the mediator field, from the numerical integration of  Eq.~(\ref{signal_nl_dynamics}),  as a function of space and time. This simulation was done assuming a fast dynamics, $\tau_\phi = 1/500$ with $\gamma=\mu=D_\phi/4=1$, $D_\phi/4$, $\nu=4$ and $\mu=1$. Pulse duration was $w_\xi = 0.01$ and amplitude $I_0=10^2/\tau_\phi$. See~\ref{ap:numerical} for details. In the right panel, the solid black line shows the predicted steady profile, given by Eq.~(\ref{kernel}), and the symbols are the result of the numerical simulation in the left panel, taken at the time indicated by the vertical dashed line, for which the mediator field is close to achieving its steady configuration.  For comparison, the right panel shows some other kernels $\cG_{\mu\nu}$ from Eq. (\ref{kernel}), where the subindices indicate the values of $\mu$ and $\nu$ used. They are the exponential kernel, $\mathcal{G}_{1,1}$, as a solid gray line, and the triangular case, $\mathcal{G}_{1,3}$, as the dashed black line. These last kernels are  normalized (indicated by $\dagger$) to match the peak of the $\cG_{1,4}$ case.}
        \label{fig:pulse}
    \end{figure}
\end{center}

Our choice of mediator equation has the advantage that some analytical results are available: For the fast mediator-dynamics case,  Eq. (\ref{G}), which now reads
\BE
D_\phi\partial_x\left(\cG^{\nu-1}\partial_x \cG\right) - \gamma \cG^\mu+ I_0 \delta(x) = 0 \ ,
\EE
can be exactly solved analytically \cite{Tsallis1996} (see also the Supplemental Material in \cite{Colombo2023} for a detailed calculation), yielding
\begin{align}
\label{kernel}
\mathcal{G}(x) &= A
\left[ 1 - (1-q) \left| s x \right| \right]^{\frac{1}{1-q}}\, ,
\ A=\left[ \frac{I_0}{2 D_\phi}\sqrt{\frac{\mu+\nu}{2\kappa}} \right]^{\frac{2}{\mu+\nu}}\ ,
\end{align}
with $q = 1+(\mu-\nu)/2$, $s^2 = 2 \kappa A^{\mu-\nu}/(\mu +
\nu)$, and $\kappa=\gamma/D_\phi$. If $q < 1$ the support of
this solution is restricted to $|sx|\leq 1/(1-q)$. The kernel $\mathcal{G}$ can range from smooth exponential profiles (for $\mu=\nu$), to truncated profiles (e.g. the triangular, $\nu-\mu=2$). These cases and additional examples are provided in Fig.~\ref{fig:pulse}.

We now consider two different ways in which the population and the signal or mediator interact: a population that releases a toxin to the medium, and a signal release that affects population mobility.

\subsection{Toxic pulses}
\label{sec:reaction}
For the first example, we revisit the model previously considered in~\cite{Colombo2023}.
There, the individuals of the focal population
release a toxic substance that reduces reproductive success within a given neighborhood.
Such a mechanism works to secure
resources or to create a safer surrounding environment.
In this case, our model for the population dynamics follows
\BE
\label{populationdyn}
\cL[\rho] = (D_\rho\partial_{xx} + r)\rho \, \, , \, \, \, \,
\cF[\rho,\phi] = - \epsilon \phi \rho \ ,
\EE
so that Eq. (\ref{geneq1}) reads
\begin{align}
\label{toxin}
\tau_\rho\partial_t \rho = (D_\rho\partial_{xx} + r)\rho - \epsilon \phi \rho \ ,
\end{align}
which includes simple diffusion with diffusion coefficient $D_\rho$ arising from Brownian motion of the organisms,
and population growth with
rate $r$, indicating the difference between reproduction and death rates. This choice is the basis for more complex models~\cite{murray2002}. $\epsilon$ is an exposition factor giving the population sensitivity to the harmful substance $\phi$. It should be mentioned that we have factorized the timescales $\tau_\rho$ and $\tau_\phi$ in Eqs. (\ref{toxin}) and (\ref{specificeq2}), therefore, the actual physical parameters (with correct units) are $D_\rho/\tau_\rho$, $\gamma/\tau_\phi$, for example. For simplification, through the rest of the paper, we will conveniently refer to the parameters ignoring this factorization (e.g. calling $D_\rho$ the diffusion coefficient).

To complete the model we should specify the rate for pulse release. It is enough for our purposes to assume here that pulse releases from the different individuals are independent events occurring at a fixed rate $\alpha$, from which the total rate at a space-time point with density $\rho$ will be $H(\rho) = \alpha\rho$. The temporal rate $R_H(t)$ is now proportional to the total population at time $t$: $R_H(t)=\alpha\int dx \rho(x,t)$.

We consider first the slow-mediator limit, for which Eqs. (\ref{sloweqs}) now become
\begin{align}
\label{ToxinSloweqs}
\notag
\tau_\rho\partial_t \rho &= (D_\rho\partial_{xx} + r)\rho - \epsilon \phi \rho\,  ,\\
\tau_\phi\partial_t \phi &= D_\phi\partial_x(\phi^{\nu-1}\partial_x \phi) - \gamma \phi^\mu + \hat I \rho  \ ,
\end{align}
with $\hat I \equiv  \alpha w_\xi I_0$. The steady and homogeneous solutions, $\rho_0$ and $\phi_0$, to this system can be found by setting all spatial and temporal derivatives to zero. Besides the trivial $\rho_0=\phi_0=0$ solution, we find $\rho_0=(\gamma/\hat I)(r/\epsilon)^\mu$, $\phi_0=r/\epsilon$. We can test the stability of this homogeneous solution to pattern formation by introducing perturbations $\rho(x,t)=\rho_0+\delta\rho(x,t)$, $\phi(x,t)=\phi_0+\delta\phi(x,t)$ and linearzing Eqs. (\ref{ToxinSloweqs}). Then, we solve the resulting equations for monochromatic perturbations, i.e. $(\delta\rho,\delta\phi)=(u,v) e^{\lambda(k)t} e^{ikx}$. The resulting growth rates are
\begin{align}
\label{lambdatoxinslow}
\lambda(k)_\pm = \frac{1}{2}\left(-b\pm\sqrt{b^2-4c}\right)\ ,\, \, \,
&\textrm{with} \ \
b=\frac{D_\rho}{\tau_\rho} k^2 +\frac{1}{\tau_\phi}\left( D_\phi\phi_0^{\nu-1}k^2+\gamma\mu\phi_0^{\mu-1} \right)
\notag \\
&\textrm{and}  \ \ c= \frac{1}{\tau_\rho \tau_\phi} \left( \epsilon\rho_0 + D_\rho k^2 \left( D_\phi\phi_0^{\nu-1}k^2+\gamma\mu\phi_0^{\mu-1} \right)\right)
\ .
\end{align}
As we take all parameters in the model to be positive, $c$ is also positive, from which we have that the growth rates are always negative. Thus, in the slow-mediator limit of Eqs. (\ref{ToxinSloweqs}), there is no pattern-forming instability.

\begin{center}
    \begin{figure}
        \raggedleft
        \hspace{1.5cm}\includegraphics[width=0.9\columnwidth]{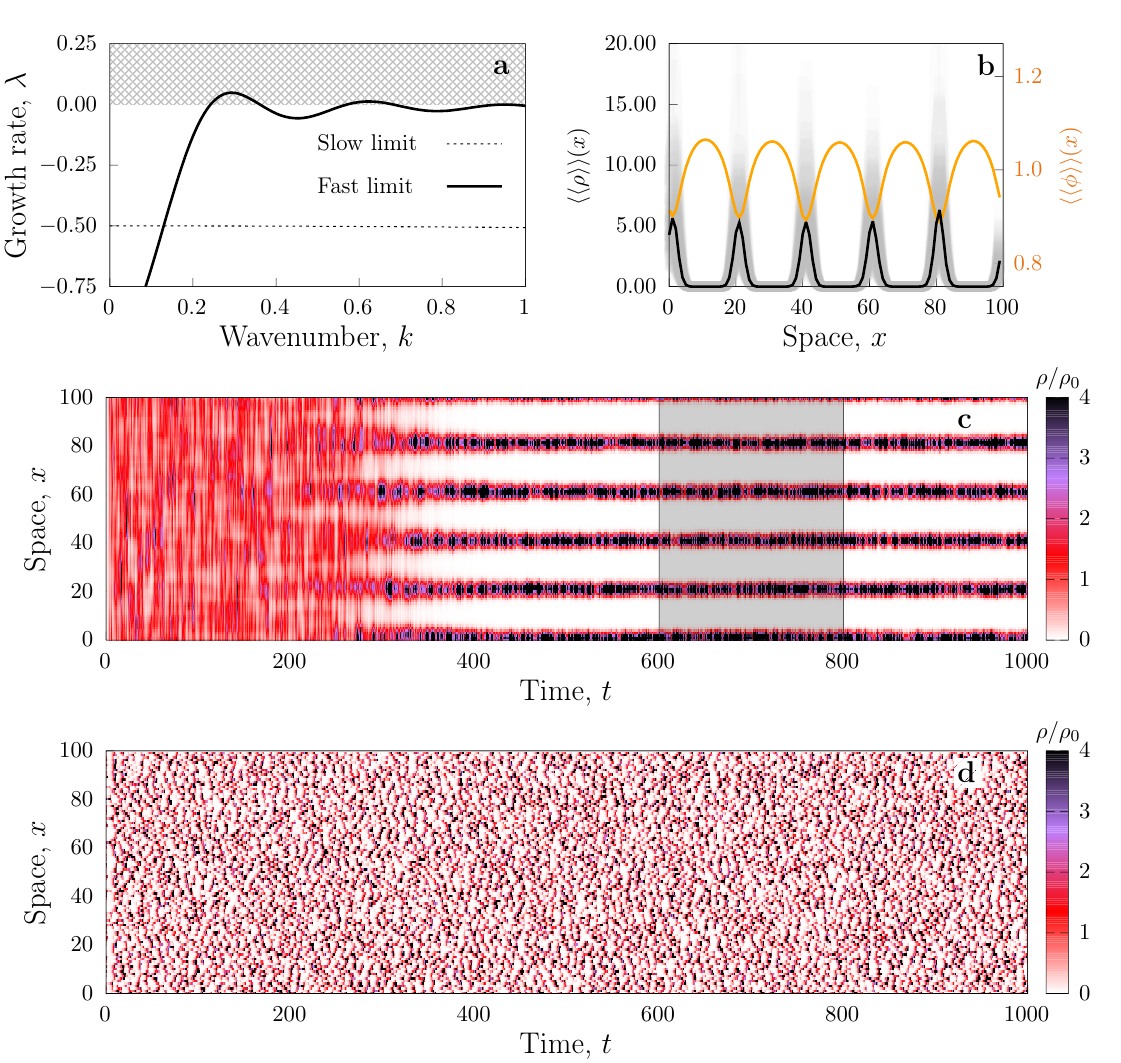}
        \caption{\textbf{Pattern formation induced by toxic pulses}. a) Growth rate ($\lambda_+(k)$, which is the largest branch) for the slow-mediator limit ($\tau_\phi/\tau_\rho = 1$, Eq. (\ref{lambdatoxinslow})), and growth rate $\lambda(k)$ for the fast-mediator ($\tau_\phi/\tau_\rho \to 0$, Eq. (\ref{lambdatoxinfast})) limit. The curve `slow limit' looks flat but actually decays following approximately $-k^2$ at large $k$.
        The gray region highlights the instability condition, $\lambda>0$. b) Long-time patterns for the fast mediator-dynamics limit. Black and orange solid lines represent, respectively, the time-averaged density distribution for the population and mediator obtained from simulations (averaged over the gray time-window indicated in panel (c), namely $t\in [600,800]$, i.e. $\langle \langle \rho(x) \rangle\rangle = (1/200)\int_{600}^{800} \rho(x,t) dt$, same for $\phi$). In the background, gray solid lines display population densities observed in the simulation in panel (c) at many times during the same time-window. Panel (c) shows the spatiotemporal evolution of the population density for a fast-mediator limit ($\tau_\rho/\tau_\phi = 500$), while panel (d) shows a slow-mediator ($\tau_\rho/\tau_\phi = 1$)  case. Colors represent the scaled density $\rho/\rho_0$ as indicated.
        The population-dynamic parameters used were $D_\rho=10^{-2}$, $r=\tau_\rho=1$, $\alpha=10^2$, $w_\xi = 10^{-2}$, $\epsilon=10$, while the parameters governing the mediator were $D_\phi/4 = \gamma = 1$, $I_0 = 10^{2}$, $\mu=1$, $\nu=4$ and timescale, $\tau_\phi$, was $1/500$ in c), and $1$ in d). See~\ref{ap:numerical} for simulation details.}
        \label{fig:linear_toxic}
    \end{figure}
\end{center}

We now consider the fast-mediator limit of this population-toxin model, Eqs. (\ref{toxin}), (\ref{specificeq2}) and $H(\rho)=\alpha\rho$. The toxin field $\phi$ retains its stochastic and pulsing character, and is given by Eq. (\ref{eq:phipulsed}). However, using (\ref{populationdyn}), the general result in Eq. (\ref{fasteq}) for the population becomes the deterministic equation:
\BE
\label{FKKP}
\tau_\rho\partial_t \rho(x,t) = (D_\rho\partial_{xx} + r)\rho(x,t) - \bar\epsilon \rho(x,t)\int dx' \cG(x-x')\rho(x',t) \ ,
\EE
with $\bar\epsilon=\alpha w_\xi \epsilon$. This recovers the result of \cite{Colombo2023} in the framework of the more general formalism developed in Sect. \ref{sec:fastmediator}. Equation (\ref{FKKP}) is the well-studied nonlocal Fisher-Kolmogorov equation which is known to undergo pattern formation for some kernels $\cG$ \cite{Lopez2004,Fuentes2004}. To see this, it is convenient to introduce the Fourier transform of the kernel as $\tilde\cG(k)\equiv\int \cG(x) e^{ikx} dx$. The steady and homogeneous non-vanishing solution to Eq. (\ref{FKKP}) is $\rho_0=r/(\bar\epsilon\tilde\cG(0))$. As before, we can test the linear stability of this solution by introducing perturbations as $\rho=\rho_0+\delta\rho$ and finding the growth rate of perturbations of wavenumber $k$. The result is
\BE
\label{lambdatoxinfast}
\lambda(k) = -D_\rho k^2 - r \frac{\tilde\cG(k)}{\tilde\cG(0)}\ .
\EE
This can
become positive for some $k$ if the Fourier transform $\tilde\cG(k)$ of the kernel takes negative values. Specifically, instability occurs if the population diffusion coefficient is sufficiently small and the
interaction kernel in Eq. (\ref{G}) sufficiently compact,
which happens when $q=1+(\mu-\nu)/2<1$ (sub-exponential). Figure
\ref{fig:linear_toxic}a) displays the growth rate in this case, compared with an example of Eq. (\ref{lambdatoxinslow}) for a value of $\tau_\phi$ more appropriate for the slow-mediator limit.

We have also performed direct numerical simulations of the full population-toxin model, Eqs. (\ref{toxin}) and (\ref{specificeq2}). The numerical algorithm is detailed in~\ref{ap:numerical}. Results are also presented in Fig. \ref{fig:linear_toxic}. As predicted from the theoretical framework, the population remains unstructured when the parameter timescales are close to the slow-mediator dynamics limit, and pattern formation occurs in the fast-mediator dynamics limit.
Naively one could expect that toxin concentrations are higher where population density is large, since the toxin is produced by the population. Nevertheless, the orange line in Fig. \ref{fig:linear_toxic} shows that exactly the contrary occurs. In fact, the presence of periodically spaced locations with excess toxin explains why the population nearly vanishes there and accumulates between them. It would be difficult to understand this from the original model Eqs. (\ref{toxin}) and (\ref{specificeq2}), but the mechanism has been already elucidated for nonlocal models such as Eq. (\ref{FKKP})
\cite{Martinez-Garcia2013,Martinez-Garcia2023}: Toxin is mainly released at the population peaks but rapidly diffuses from there according to the nonlinear diffusion Eq. (\ref{specificeq2}). Although in the fast-mediator limit different pulses will not overlap, in the coarse-grained effective description given by Eq. (\ref{meanphi}) with $H(\rho)=\alpha\rho$, places in-between population peaks receive toxin from the two neighboring peaks and then, for a class of kernels $\cG$, toxin concentration there can exceed the levels close to each of the individual peaks. This enhancement mechanism becomes optimal when the distance between population peaks is between 1 and 2 times the interaction range defined by $\cG$ \cite{Martinez-Garcia2013,Martinez-Garcia2023}.

\begin{center}
    \begin{figure}[h]
        \centering\hspace{2cm}     \includegraphics[width=0.6\columnwidth]{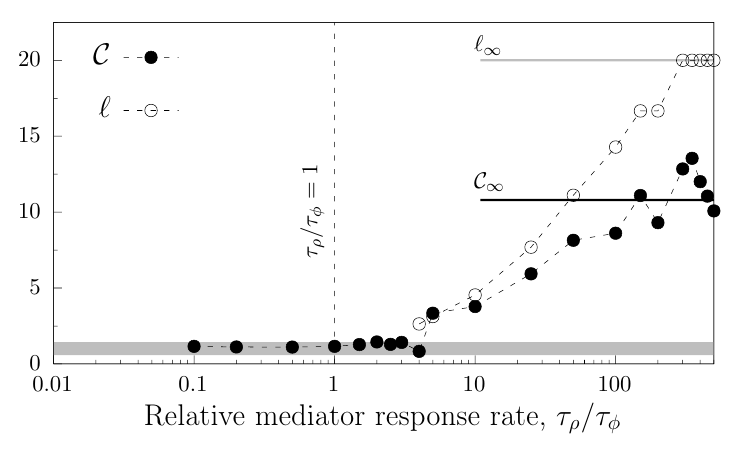}
        \caption{\textbf{Timescale transition}. Pattern coherence, $\mathcal{C}$, and wavelength, $\ell$, as a function of $\tau_\rho/\tau_\phi$, which is the ratio of the mediator response rate, $1/\tau_\phi$, to the population response rate $1/\tau_\rho$. Coherence is defined as $S(k^\star)/\bar S$, where $S$ is the time-averaged power-spectrum of the population, $k^\star$ is the mode with maximum amplitude and $\bar S$ is the average mode amplitude. The coherence and wavelength predicted by the fast mediator-dynamics limit, $\mathcal{C}_\infty$ and $\ell_\infty$, respectively, are shown as horizontal solid black and gray lines. The simulation was done with the same parameters as in Fig.~\ref{fig:linear_toxic} with the exception of $\tau_\phi$, which was varied.  A vertical dashed line highlights the slow-mediator case  ($\tau_\rho/\tau_\phi = 1$) for which our theoretical results predicts that spatial patterns can not appear. For simulation details and values of the parameters used see caption of~Fig.~\ref{fig:linear_toxic}.}
        \label{fig:timescale}
    \end{figure}
\end{center}

After the detailed theoretical analysis of the extreme timescale limits, we hypothesize that there exists an
intermediate critical timescale $\tau_\phi^c$
separating the pattern and no-pattern phases. To check
whether this critical timescale exists, we
numerically integrate Eqs.~(\ref{toxin}) and (\ref{specificeq2})
for different values of $\tau_\phi$ and
characterize the long-time patterns. Fig.~\ref{fig:timescale}
provides insights into how this transition
occurs by measuring the coherence, $\mathcal{C}$ (see definition in the caption),
and wavelength of the patterns, $\ell$, as a function
of  $\tau_\rho/\tau_\phi$. The result confirms the intuition
that there exists a critical value, $(\tau_\rho/\tau_\phi)^c$,
above which periodic patterns appear. Furthermore, the coherence and wavelength of the pattern at large values of $\tau_\rho/\tau_\phi$ stabilizes at the values produced by the long-time distribution obtained from the numerical integration of the fast limit approximation given by Eq. (\ref{FKKP}), $\mathcal{C}_\infty$ and $\ell_\infty$, respectively (see Fig.~\ref{fig:timescale}). Additionally, $\ell_\infty$ can also be predicted by the fastest growing mode (the wavelength associated to the wavenumber of the maximum of $\lambda(k)$). Below the critical value,
the long-time population density
is statistically homogeneous (with unstructured  fluctuations).

\subsection{Motility-inducing pulses}
\label{sec:motility}

The previous case addressed the role of pulsed
interactions on pattern formation when mediators
influence a reaction process~\cite{Colombo2023}. 
In this section, we will take a first step extending our framework to motility-controlling pulses by focusing on a minimal model in which the mediating substance affects motility through a dependence in the diffusion coefficient, $D_\rho=D_\rho(\phi)$:
\BE
\mathcal{L}[\rho]=0 \ \ , \ \ \ \cF[\rho, \phi] = \partial_{xx}( D_\rho(\phi)\rho) \ .
\label{LFmotility}
\EE
Later we will focus on the case in which $D'_\rho(\phi)>0$, i.e. the mediator signal increases diffusivity.
Under this scenario, the population dynamics reduces
to a diffusion equation of a form related to ~\cite{clopez2006}:
\begin{align}
\label{motility}
\tau_\rho \partial_t\rho &= \partial_{xx}( D_\rho(\phi)\rho) \, .
\end{align}
Note that in the general development of Sect. \ref{sec:fastmediator} we assumed that $\cF[\rho, \phi=0]=0$. This implies that population diffusion vanishes in the absence of signal, $D_\rho(0)=0$. We will consider this simplification but it is worth mentioning that it is straightforward to extend the calculation to cases in which $D_\rho(0)\neq 0$ by substituting $D(\phi)$ by $D_\rho(\phi)-D_\rho(0)$, and taking $\mathcal{L}[\rho]=D_\rho(0)\partial_{xx}\rho$.
We also note that, as mentioned before, we are abusing the language when calling $D_\rho(\phi)$ a diffusion coefficient. What has really the meaning and dimensions of a diffusion coefficient is $D_\rho(\phi)/\tau_\rho$.  The mediator $\phi$ is a kind of biochemical signal released by the individuals, and that stimulates the mobility of conspecifics. $\phi$ is released in the form of pulses $\xi_\rho$, undergoing the dynamics of Eq. (\ref{specificeq2}) before feeding back at each location to the value of the population diffusion coefficient there, $D(\phi(x,t))$.

Eq.~(\ref{motility}) is the macroscopic description
of a state-dependent random walk under
the Ito prescription. Here we adopted the
Ito prescription, instead of Stratonovich or kinetic,
because it is a non-anticipative
prescription that reflects an intrinsic
randomness of individuals
decision-making process~\cite{hanggi1994colored,clopez2006,dos2020critical}.

To completely define the model, we need to specify the spatiotemporal release rate $H(\rho)$ (further specification of the function $D(\rho)$ will be done later). We consider that individuals
found in high population density regions will
emit the mediator signal more frequently than
those in low density regions,
mimicking a quorum-sensing behavior~\cite{miller2001quorum}.
For this purpose, we
take $H(\rho)=\alpha(\rho/\bar\rho)^\beta$, with $\beta>1$.  $\bar\rho$ is a reference density value.

As before, we first consider the slow-mediator dynamics limit. In this limit the model reduces to
\begin{align}
\label{MobSloweqs}
\notag
\tau_\rho\partial_t\rho &= \partial_{xx}( D_\rho(\phi)\rho) \, \,  ,\\
\tau_\phi\partial_t \phi &= D_\phi\partial_x(\phi^{\nu-1}\partial_x \phi) - \gamma \phi^\mu + \tilde I \rho^\beta  \,
\end{align}
with $\tilde I= \alpha I_0 w_\xi/\bar\rho^\beta$.

Any positive constant $\rho_0$ is a steady and homogeneous solution of (\ref{MobSloweqs}) provided the corresponding value of $\phi_0$ satisfies $\gamma \phi_0^\mu=\tilde I \rho_0^\beta$. We test the stability of this homogeneous solution by introducing perturbations $\rho=\rho_0+\delta\rho$ and $\phi=\phi_0+\delta\phi$, assumed to be monochromatic of wavenumber $k$, and then study the short-time dynamics at first-order in $\delta \rho$ and $\delta \phi$. The corresponding growth rates just after the perturbations are given by
\begin{align}
\label{lambdamobslow}
\lambda(k)_\pm &= \frac{1}{2}\left(-b\pm\sqrt{b^2-4c}\right)\ , \notag \\
&\textrm{with} \ \
b=\frac{D_\rho(\phi_0)}{\tau_\rho} k^2 +\frac{1}{\tau_\phi}\left( D_\phi\phi_0^{\nu-1}k^2+\gamma\mu\phi_0^{\mu-1} \right)
\notag \\
&\textrm{and}  \ \ c= \frac{1}{\tau_\rho \tau_\phi} \left( \gamma\beta\phi_0^\mu D'_\rho(\phi_0)k^2 + D_\rho(\phi_0) k^2 \left( D_\phi\phi_0^{\nu-1}k^2+\gamma\mu\phi_0^{\mu-1} \right)\right)
\ .
\end{align}
$D_\rho(\phi_0)$ and $D'_\rho(\phi_0)$ are, respectively, the value of the population diffusion coefficient and its derivative, evaluated at the signal homogeneous value $\phi_0$. We see that as long as $D'_\rho(\phi_0)>0$, we have $c>0$, from which the growth rates remain negative for any (positive) value of the other parameters. No pattern formation occurs under these conditions. We thus restrict to $D_\rho(\phi)$ increasing with the signal $\phi$ to be sure that any pattern-forming instability that occurs in model (\ref{motility})-(\ref{specificeq2}) (with $H(\rho)=\alpha(\rho/\bar\rho)^\beta$), beyond the slow-mediator limit, would originate from the pulsed nature of signal releases.

We analyze now the limit of fast-mediator dynamics. The mediator remains pulsed and stochastic, as given by (\ref{eq:phipulsed}), but the population is described by the effective equation (\ref{fasteq}). Using (\ref{LFmotility}), it reads
\begin{align}
\label{effectivemot}
\tau_\rho\partial_t\rho(x,t) &= \alpha w_\xi \partial_{xx}\left( \rho(x,t)\int D_\rho\left( \cG(x-x')\right) \left(\frac{\rho(x',t)}{\bar\rho}\right)^\beta dx'\right)  \notag \\
&= K \partial_{xx} \left( \rho(x,t)\int G(x-x') \rho(x',t)^\beta dx'\right) \ ,
\end{align}
where we have defined $K \equiv \alpha \omega_\xi / \bar\rho^\beta$ and an effective kernel that takes into account both the mediator steady-pulse shape $\cG$ and the mediator-dependent diffusion coefficient: $G(x) \equiv D_\rho\left( \cG(x)\right)$. Note that, since $D_\rho(0)=0$, if $\cG$ has compact support the same happens for $G$.

\begin{center}
    \begin{figure}
        \raggedleft
        \includegraphics[width=0.9\columnwidth]{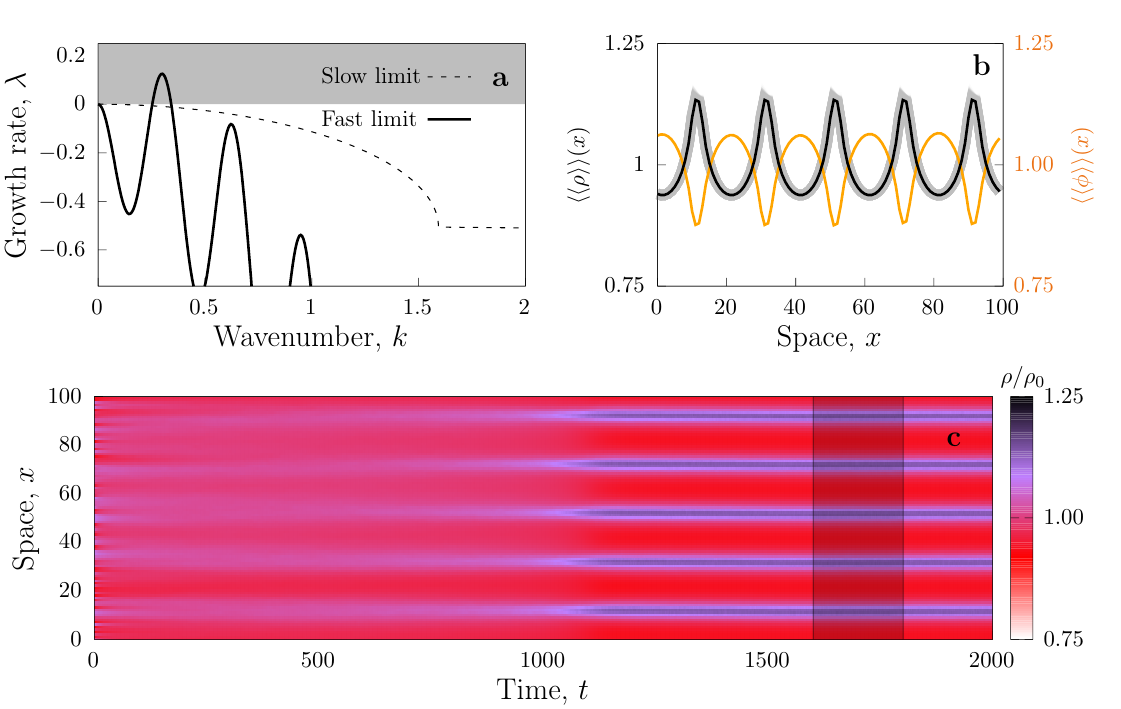}
        \caption{\textbf{Pattern formation emerging from motility-inducing pulses}. a) Growth rate ($\lambda_+(k)$, which is the largest branch) for the slow-mediator limit ($\tau_\phi/\tau_\rho = 1$, Eq. (\ref{lambdamobslow})), and growth rate $\lambda(k)$ for the fast-mediator ($\tau_\phi/\tau_\rho \to 0$, Eq. (\ref{lambdamobfast})) limit. The gray region highlights the instability condition, $\lambda>0$. b) Long-time patterns for the fast mediator-dynamics limit. The black and orange solid lines represent, respectively, the time-averaged density for the population and mediator obtained from simulations (averaged over the gray time-window indicated in panel (c), namely $t\in [600,800]$, i.e. $\langle \langle \rho(x) \rangle\rangle = (1/200)\int_{1600}^{1800} \rho(x,t) dt$, same for $\phi$). In the background, gray solid lines display population densities observed in the simulation in panel (c) at many times during the same time-window. Panel (c) shows the spatiotemporal evolution of the population density for a fast-mediator case ($\tau_\phi/\tau_\rho \to 0$ with $\tau_\rho=1$, see~\ref{ap:numerical} for simulation details). Colors represent the scaled density $\rho/\rho_0$ as indicated.
        Parameters used for the population dynamics were $D_\rho(\phi)=\phi$, $r=\tau_\rho=\rho_0=\bar \rho = 1$, $\alpha=10^{-1}$, $w_\xi = 10^{-2}$ and $H(\rho)=\alpha\rho^{\beta}$ with $\beta=50$. The mediator's parameters were $D_\phi/4 = \gamma = 1$, $I_0 = 10^{2}$, $\mu=1$, $\nu=4$.}
        \label{fig:linear_mot}
    \end{figure}
\end{center}

Any positive constant $\rho_0$ gives a steady and homogeneous solution to Eq. (\ref{effectivemot}). As before we can check its stability to perturbations of wavenumber $k$ by introducing $\rho=\rho_0+\delta\rho$ and linearizing. In terms of the Fourier transform $\tilde G(k)=\int G(x) e^{ikx} dx$, the growth rate of perturbations of wavenumber $k$ is
\BE
\label{lambdamobfast}
\tau_\rho\lambda(k) = - D_0 k^2\left( 1+\beta\frac{\tilde G(k)}{\tilde G(0)}\right) \  ,
\EE
with $D_0\equiv  K \tilde G(0) \rho_0^\beta$. Thus, we see that if $\tilde G(k)$ takes negative values for some $k$, $\lambda(k)$ would become positive for sufficiently large $\beta$, indicating the instability of the homogeneous $\rho_0$ to pattern formation.
Panel a) of Fig.~\ref{fig:linear_mot} displays the linear growth rates of perturbations in the slow- and fast-mediator limits.

We have also performed numerical simulations of Eqs. (\ref{motility}) and (\ref{specificeq2}) (see Appendix \ref{ap:numerical} for numerical details), considering a linear diffusion coefficient, $D(\phi)=\phi$ (which implies $G=\cG$). Initial condition is a constant $\rho_0$ with small random fluctuations. In the slow-mediator case, $\rho(x,t)$ and $\phi(x,t)$ remain statistically homogeneous, with unstructured fluctuations (not shown). But for sufficiently small $\tau_\phi$ (fast-mediator case), pattern formation occurs. Figure \ref{fig:linear_mot} shows the pattern-forming process (panel c) and the statistically steady state for both $\rho$ and $\phi$ (panel b). We also see here how the population avoids the accumulation zones of $\phi$, since the enhanced mobility there makes the field $\rho(x,t)$ to remain there for less time. And on the other hand the maxima of $\phi$ occur in between population peaks, since the mediator arrives there from two different places of strong production.  We see that this motility case is another situation in which the pulsed character of the releases changes qualitatively the stability of the homogeneous distributions compared to the case in which only the constant average of the pulses is introduced into the system.

\section{Final remarks and perspectives}

We have introduced a Pulsed Interaction Framework consisting in a population that releases a mediator in the form of pulses. The mediator spatially propagates and evolves, and then the population is affected by the distribution of this mediator field. In a slow-mediator limit, properly coarse-graining the dynamics results in a reaction-diffusion system in which the mediator's pulsed release is effectively replaced by its average behavior. In the fast-mediator limit, however, we obtain a nonlocal model in which the long-range interaction kernel can be derived from the mediator dynamics. Therefore, by only changing one single control parameter, the mediator timescale, our framework unifies hitherto complementary approaches to modeling spatial population dynamics \cite{murray2003}. We have analyzed two examples in which the change of the structure of the effective models for the two limits reveals that pattern formation occurs in the presence of pulsed interactions, but becomes impossible if only the mean value of the pulses is considered. In addition, the effective descriptions clarify the mechanisms responsible for the observed outcomes. Overall, as we reveal this critical role of the mediator dynamics and timescale, we highlight that the details of how interaction is established should not be overlooked when developing biological or ecological models, especially because interactions mediated by pulsed signals are not rare in nature, being observed in cases from microbes to larger organisms.

Organisms release pulses of different mediators to establish different kinds of interactions, which, importantly, are very often responsible for the emergence of self-organized population-level spatiotemporal patterns. For example, starving cells of the social amoeba \textit{Dictyostelium discoideum} use chemical pulses to synchronize their behavior and initiate an aggregation process that culminates with the formation of multicellular fruiting bodies in which some cells increase their dispersal ability \cite{Gregor2010a,Devreotes1983}.
Similarly, flashing electrochemical communication underpins the behavioral synchronization of bacteria~\cite{ojalvo2018}, allowing the coordination of actions at distance, and of male fireflies during courtship~\cite{sarfati2021self}.
At much larger scales, several bird species sing to defend their breeding and non-breeding territories (i.e., pulsed acoustic signals) \cite{Collins2004,Scheffer2006}. Ungulates like the Mongolian gazelle emit loud calls that can impact their foraging efficiency \cite{ricardo2013p}.
Finally, Meerkats perform vocal exchanges that cumulatively result in decision-making processes relative to group movement \cite{demartsev2018vocal}.
Although our framework is not specialized on any of these cases, it provides the right mathematical formalism for future studies to investigate the interplay between the spatiotemporal statistics of signaling processes and emergent spatiotemporal patterns.

Future work should account for the fact that, as the emergent spatial patterns can affect organisms' fitness, the mediator dynamics and the timescales involved are likely to be influenced by spatial eco-evolutionary feedbacks~\cite{colombo2019spatial,Nadell2016,Nowak2006,Lieberman2005}. For example, grouping allows populations subjected to an Allee effect (inverse density dependence), such as some fish, rotifer, or mammals, to persist in harsh environments where they would go extinct if uniformly distributed \cite{Jorge2023}. In these species, aggregation is beneficial because cooperation and facilitation mechanisms increase the probability of survival and/or reproduction ~\cite{Allee1932,Stephens2002,Ghazoul2005}. In other cases, the segregation of individuals maximizes individuals' survival probability, for example, by avoiding competition for resources~\cite{connell1963territorial}. In general, the positive and negative effects of aggregation are typically present concomitantly, and their interplay is responsible for population spatial self-organization.

Finally, validating any of these possible theoretical
extensions as well as our results regarding the interplay
between interaction signals and species' spatial organization
requires a closer interaction with empirical data. For
instance, natural environments are noisy and complex, having
many species and mediators at work. Therefore, assessing
whether the assumptions in our calculation hold in realistic
scenarios, especially the non-overlapping condition
(Eq.~\ref{eq:phipulsed}), must be investigated case by case. On
the other hand, the empirical data must correlate the positions
of biological entities with the dynamics of the associated
mediator. Such approaches have recently been made possible both
in laboratory
settings~\cite{curatolo2020cooperative,ojalvo2018} and in the
field~\cite{demartsev2023signalling}, which create
opportunities to properly parameterize our framework for
specific biological systems, provide mechanistic explanations
for the observed phenomena, and make new theoretical
predictions to be tested with novel experimental designs.

\vspace{1cm} \textit{Acknowledgements.---} This work was
partially funded by the Center of Advanced Systems
Understanding (CASUS) which is financed by Germany's Federal
Ministry of Education and Research (BMBF), and by the Severo
Ochoa and Maria de Maeztu Program for Centers and Units of
Excellence in R\&D, grant MDM-2017-0711 funded by
MCIN/AEI/10.13039/501100011033. C.L. and E.H.-G. also
acknowledge grant LAMARCA PID2021-123352OB-C32  funded by
MCIN/AEI/10.13039/501100011033 and FEDER ``Una manera de hacer
Europa''. J.M.C. was supported by NSF IIBR 1915347.

\section*{References}
\bibliographystyle{iopart-num}
\bibliography{references_pulsed}

\appendix

\section{Numerical integration scheme}
\label{ap:numerical}
In order to numerically integrate the pulsed interaction framework (Eq.~\ref{geneq1}-\ref{releases}),
we follow a standard forward first-order Euler scheme. Space is discretized in small cells of size
$\delta x$ and time-step size is given $\delta t$, yielding the representation, $\rho_{j,n}\equiv \rho(x=j\delta x, t= n \delta t)$,
and analogously for $\phi_{j,n}$.
\subsection*{Toxic pulses}
For the toxic pulses cases defined in Sec.~\ref{sec:reaction}, the population and mediator density changes after one
time-step $\delta t$ are given by
\begin{align}
\label{eq:numerical-rho}
\rho_{j,n+1} &= \rho_{j,n} + [D_\rho(\rho_{j+1,n}+\rho_{j-1,n}-2\rho_{j,n})/(\delta x^2) +
r\rho_{j,n}-\epsilon \rho_{j,n}\phi_{j,n}] \delta t/\tau_\rho\, ,\\
\phi_{j,n+1} &= \phi_{j,n} + [D_\phi(\phi_{j+1,n}^\nu+\phi_{j-1,n}^\nu-2\phi_{j,n}^\nu)/(\nu\delta x^2)
- \gamma\phi_{j,n}^{\mu} + (\xi_\rho)_{j,n}] \delta t/\tau_\phi\, .  \label{eq:numerical-phi}
\end{align}
where we used that the nonlinear diffusion term
$\partial_x(\phi^{\nu-1}\partial_{x}\phi)$ can be written as
$\nu^{-1}\partial_{xx}\phi^\nu$ to mitigate possible numerical
instabilities. The stochastic pulses, represented by $(\xi_\rho)_{j,n}$,
are implemented as follows: Initially all $(\xi_\rho)_{j,n}$
are set to zero. At each time step, a pulse can occur somewhere in the system, with probability $R_H \delta t$, where $R_H=\sum_k H(\rho_{k,n})\delta x$. If a pulse occurs, its location $j$ is sampled
from the probability distribution $H(\rho_{j,n})/R_H$. Then, the
value $(\xi_\rho)_{j,n}$ is set to $I_0/\delta x$ during
$w_\xi/\delta t$ time steps, being reset to zero
afterwards (the denominator $\delta x$ arises from the
discretization of the spatial delta function). Finally, in order to ensure that
the pulses are resolved by the numerical integration we need to consider $\delta x=1.0$ much smaller than mediators' pulse reach ($\sim
10$ for the cases we investigate), and $\delta t\in[10^{-4},10^{-6}]$, much smaller than the signal
duration time $w_\xi=0.01$.

\subsection*{Motility-inducing pulses}
For the scenario defined in Sec.~\ref{sec:motility}, the population Eq.~(\ref{motility}) can be writen in its discretized version:
\begin{equation}
\label{phinumerics}
\rho_{j,n+1} = \rho_{j,n} + [(D_\rho(\phi_{j+1,n})\rho_{j+1,n}+D_\rho(\phi_{j-1,n})\rho_{j-1,n}-2D_\rho(\phi_{j,n})\rho_{j,n})/(\delta x^2)] \delta t/\tau_\rho\, .\\
\end{equation}
Because of the use of the nonlinear rate $H(\rho) \propto \rho^\beta$ for the generation of mediator pulses, the  explicit solution of Eq. (\ref{specificeq2}) in the fast-mediator situation turns out to be computationally too costly. Thus, in this case we use Eq. (\ref{eq:phipulsed}), valid in the limit  $\tau_\phi/\tau_\rho \approx 0$, as the solution of the mediator dynamics. Then we discretize space and time
\begin{equation}
    \phi_{i,n} = \sum_m \cG(i\delta x-x_m) \Pi(n\delta t-t_m)\, ,
\end{equation}
and use this expression in the numerical solution of the population Eq. (\ref{phinumerics}).

\end{document}